\documentclass[onecolumn,superscriptaddress]{revtex4}
\usepackage{amsmath,amssymb,amsfonts,color,graphicx}

\begin{document}

\title{Synthetic reverberating activity patterns embedded in networks of cortical neurons}

\author{R. Vardi}
\affiliation{Gonda Interdisciplinary Brain Research Center, and the Goodman Faculty of Life Sciences, Bar-Ilan University, Ramat-Gan 52900, Israel.}
 
\author{A. Wallach}
\affiliation{ Network Biology Research Laboratories, Technion - Israel Institute of Technology, Haifa 32000, Israel.}

\author{E. Kopelowitz}
\affiliation{Gonda Interdisciplinary Brain Research Center, and the Goodman Faculty of Life Sciences, Bar-Ilan University, Ramat-Gan 52900, Israel.}
\affiliation{Department of Physics, Bar-Ilan University, Ramat-Gan 52900, Israel.}

\author{M. Abeles}
\affiliation{Gonda Interdisciplinary Brain Research Center, and the Goodman Faculty of Life Sciences, Bar-Ilan University, Ramat-Gan 52900, Israel.}

\author{S. Marom}
\affiliation{ Network Biology Research Laboratories, Technion - Israel Institute of Technology, Haifa 32000, Israel.}

\author{I. Kanter}

\affiliation{Gonda Interdisciplinary Brain Research Center, and the Goodman Faculty of Life Sciences, Bar-Ilan University, Ramat-Gan 52900, Israel.}
\affiliation{Department of Physics, Bar-Ilan University, Ramat-Gan 52900, Israel.}

\begin{abstract}
Synthetic reverberating activity patterns are experimentally generated by stimulation of a subset of neurons embedded in a spontaneously active network of cortical cells in-vitro.  The neurons are artificially connected by means of conditional stimulation matrix, forming a synthetic local circuit with a predefined programmable connectivity and time-delays.  Possible uses of this experimental design are demonstrated, analyzing the sensitivity of these deterministic activity patterns to transmission delays and to the nature of ongoing network dynamics.
\end{abstract}

\maketitle

\section{Introduction}
 The assumption of isomorphism between behavior and temporal patterns of reverberating synchronized neural electrical activity stands at the basis of many large scale theories of the brain \cite{1,2}. Actual measurement and controlled experimental manipulation of such temporal activity patterns are associated with severe difficulties and most of what is presently known comes from theoretical and numerical studies of simplified components, which are very remote from the full dynamical richness of neural network entities \cite{3,4,5,6,7,8}.
Here, we artificially generate temporal patterns of reverberating synchronized neural electrical activity by conditioned sequential stimulation of a circuit of neurons embedded within a large scale, spontaneously active network of cortical cells in-vitro \cite{9,10,11}. Cortical neurons were obtained from newborn rats within 24 h after birth using mechanical and enzymatic procedures described in earlier studies \cite{9,10,11,12,13,14,15}. The neurons were plated directly onto substrate-integrated multi-electrode arrays (60 Ti/Au/TiN extracellular electrodes) and allowed to develop functionally and structurally mature networks over a time period of 2-3 weeks [see material and methods section].

\begin{figure*}
\begin{center}
\includegraphics[width=0.9\textwidth]{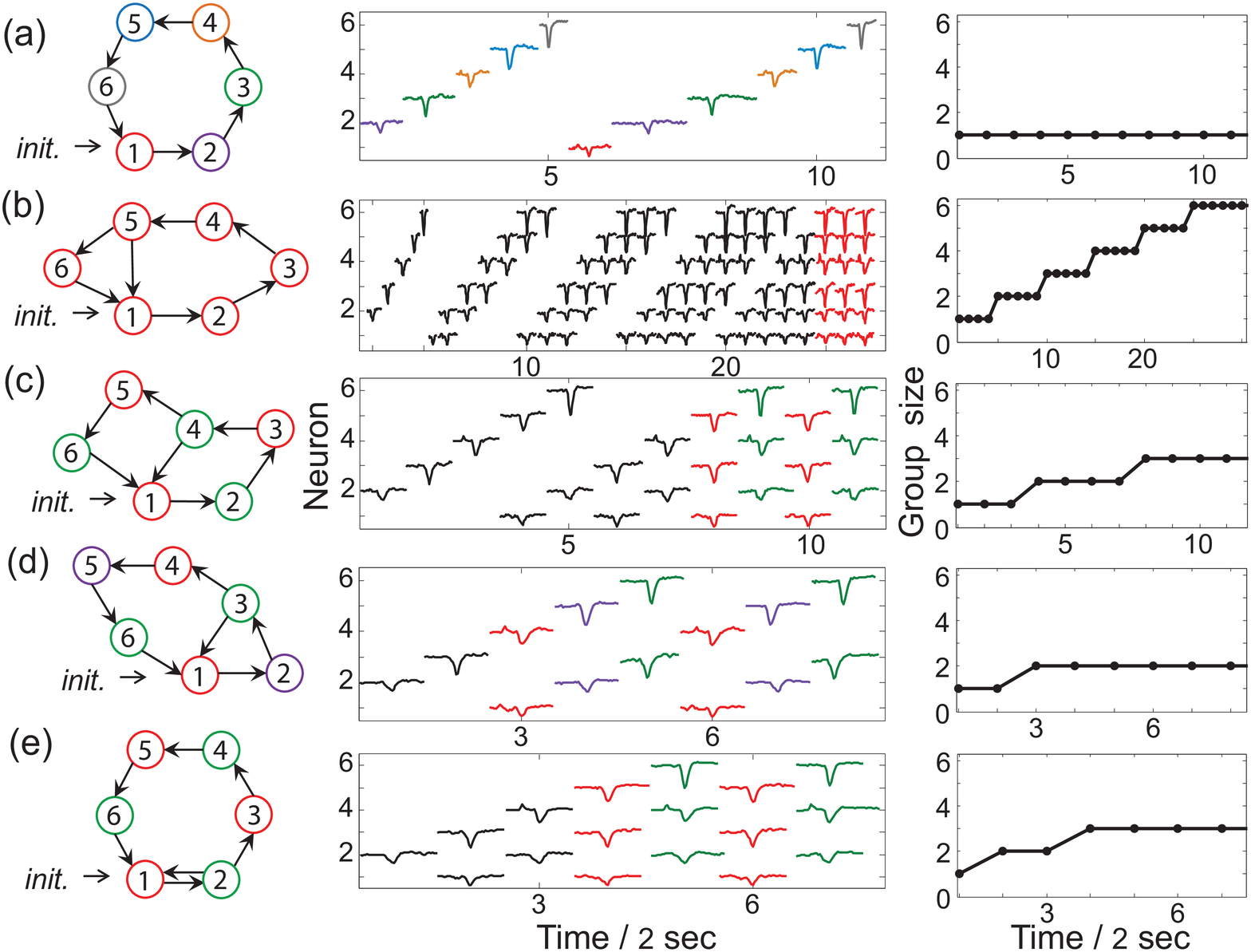}
\end{center}
\caption{(Colour on-line) Transients and reverberating activity patterns in six neuron circuits, using a conditioned stimulation protocol following the circuit connectivity (arrows in the left column). All programmable delays set equal to 2 s and at time zero neuron 1 fired. The second column shows spike trains of the six recording electrodes. In this column the abscissa shows the experimental time, but the spikes are plotted on an expanded time scale of ~2.5 ms per trace, indicating that we related to clear spikes with good signal to noise ratio and with reliably repeating shapes. Some of the neurons rarely fired also in between external stimuli (not shown). The third column sums up the number of simultaneously active electrodes.  Neurons that fire together in the reverberating phase are represented by the same colour (first column) and after the transient trajectory of activity (second column). (a) A directed loop with 6 activity groups of one neuron each, and with the lack of a transient. (b) GCD(5,6)=1 group of reverberating activity and 25 time steps transient trajectory. (c) GCD(4,6)=2 groups of reverberating activity and 9 time steps transient trajectory. (d) GCD(3,6)=3 groups of reverberating activity and 5 time steps transient trajectory. (e) A single directed loop with one bidirectional coupling forming a loop of size 2, GCD(2,6)=2 groups of reverberating activity as in (c), but with 5 time steps transient trajectory.}
\end{figure*}

\section{Experimental setup and the GCD clusters}
The basic experimental design consists of circuits with six different neurons that are artificially coupled to each other (Fig. 1 left column). The coupling is realized in the following way: Neuron 1 is stimulated electrically, if generating a spike within a neuronal response latency $\pm$2 ms the next neuron in the artificial circuit is stimulated after a fixed delay. If the next neuron fires within a neuronal response latency $\pm$2 ms its connecting neurons are stimulated, etc $\ldots$ Spikes that happened during the delay are ignored. Care is taken to use recording electrodes that exhibit well-formed identified spikes that are selectively and reliably evoked by a spatially unique stimulation source. Consider, for instance, the case of the directed circuit shown in the left column of Fig. 1a, where a stimulus is applied in a site that evokes one spike recorded in electrode number 1. Conditioned to the detection of a spike in electrode number 1, a stimulus is applied after a fixed delay to a site that evokes a spike detected in electrode number 2 and so on, following the connectivity of the circuit. In this trivial realization, it is expected that as long as each stimulus evokes a spike that is detected in the corresponding recording electrode, a chain of one spike per stimulation cycle will be formed (Fig. 1a, middle and right columns). The other rows (Fig. 1b-e) show the results obtained by different six neuron circuits. In each of these realizations, evoked patterns of activity relax to a theoretically expected periodicity that equals the greatest common divisor (GCD) of the circuit loops, where the number of neurons that fire in synchrony is exactly 6/GCD . For instance, the realization with 3 and 6 size loops (Fig. 1d) is governed by periodicity equals to GCD(3,6)=3, three activity groups of zero-lag synchrony (ZLS), in each of which two neurons fire simultaneously. The different examples of Fig. 1 demonstrate the sensitivity of the GCD, a global determinant of periodic steady-state mode of activity, to the circuit's connectivity matrix. The resulting patterns of synchronized periodic activities might seem counterintuitive, since the circuit consists solely of unidirectional connections (Fig. 1a-d) and zero-lag neurons do not share the same input as in the relay mechanism where two neurons are unidirectionally mastered by a third one. But as shown elsewhere \cite{16,17}, the shared information in such graphs is a consequence of the information mixing mechanism where the activation of a given node is determined by the activity of nodes in different time steps. The mixing occurs during the transient trajectory, and varies between zero and 25 time steps in the cases demonstrated here (Fig. 1a-b, right column).

\section{Response failure}
Under conditions where the network \textit{spontaneous} activity is made sparse using synaptic blockers, the stability of our synthetically evoked reverberating activity patterns is largely determined by the response reliability of the stimulated circuit neurons. While the response of a single cortical neuron to repeated stimuli is inherently noisy, a reliable 1:1 response over extended time scales is achievable in our setup for 1-5 Hz stimulation frequencies, whereas higher frequencies may lead to intermittent responsiveness \cite{15}. Such response fluctuations compromise the stability of activity patterns in the simple case of a chain of one spike per stimulation cycle (Fig. 1a). But in cases where the reverberating pattern relies on several neurons that fire simultaneously, the pattern is resilient to intermittent responsiveness of one neuron (Fig. 2).

\begin{figure}[h]
\begin{center}
\includegraphics[width=0.9\textwidth]{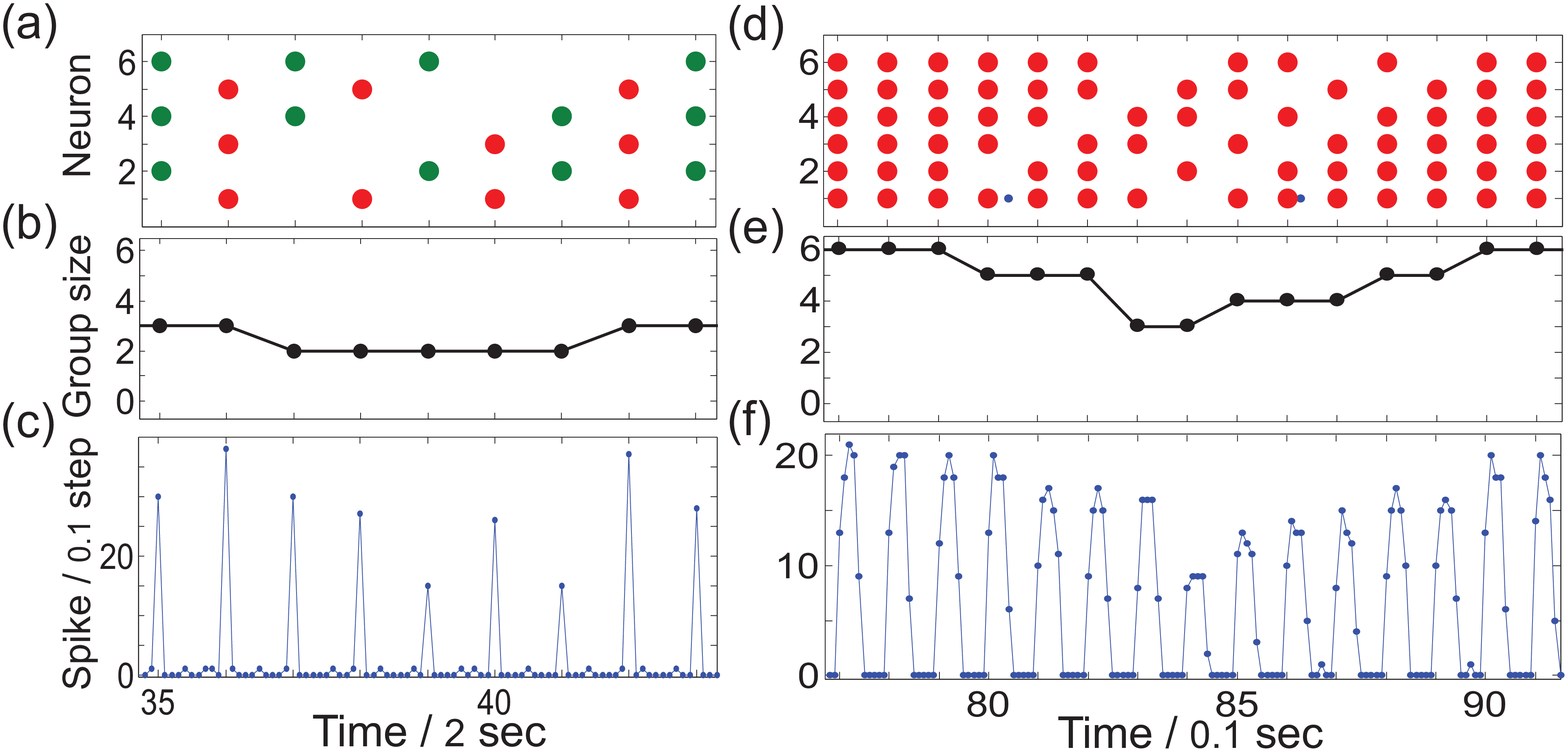}
\end{center}
\caption{ The robustness of synthetic reverberating activity patterns to transmission delay. The time of each spike is marked as a dot. Spikes that resulted due to electrical stimulation are marked by large dots and the "spontaneous" spikes by small dots. (a) Spike trains of the circuit 1c with 2 periodical modes. A response failure of the neuron labelled 2 at time step 37 is followed by 5 time steps recovery period. (b) The number of electrodes with simultaneous activity in 2a. (c) Activity of the sixty electrodes in 2a, using non-overlapping windows of 0.2 s, indicating that the network activity relaxes shortly after stimulation of the circuit. (d) Spike trains of circuit 1b with transmission delay of 0.1 s where the reverberating mode overcomes response failures of several neurons. (e) The number of electrodes with simultaneous activity in 2d. (f) Activity of the sixty electrodes in 2d, using non-overlapping windows of 0.01 s, overlaps the time scale of relaxation from the evoked network response.}
\end{figure}

One more critical factor (beyond the individual neuron response fluctuations) that destabilizes the maintenance of reverberating activity patterns is considered: Each stimulus evokes a network response that relaxes on a time scale determined by the network properties \cite{13,14}. When the time scale of stimulation overlaps the time scale of relaxation from the evoked network response, the stability of synthetic reverberating activity patterns is heavily challenged due to correlations among response failures of the six neurons that form the embedded circuit.  In Fig. 2 we demonstrate the sensitivity of synthetic reverberating activity patterns to the relations between the two time scales. Having said that, we often detected a "self-repairing" process, as the synthetic reverberating pattern overcomes response failures of several neurons through a longer transient of recovery (Fig. 2d-e). While the general trend seems lawful (the lower the GCD the stable the synthetic activity pattern), we were not able to define a simple rule that relates stability to transmission delay and network response features; the parameter space is too complex, and includes multiple spatiotemporal correlations at different levels of organization. Nevertheless, correlations between the single neuron level (Fig. 2e) and the population level (Fig. 2f) are clearly demonstrated.

\section{Spontaneous network activity}
The ongoing spontaneous activity of mature large scale cortical networks in-vitro is composed of synchronous events, decorated with asynchronous sporadic activity of individual neurons in between these synchronies \cite{10,11,13,14,18,19}. This spontaneous ongoing activity may be made more synchronous by partial blockade of inhibitory synapses \cite{13,14}, or more asynchronous by activation of cholinergic receptors \cite{18}. In what follows we examine the sensitivity of the smallest homogeneous directed circuit leading to ZLS \cite{16}, to the nature of ongoing (spontaneous) network activity using pharmacological manipulations.  The circuit consists of four electrodes with 3 and 4 directed loops, GCD(3,4)=1 (Fig. 3a), and  a stimulation to the electrode labelled 1 is leading to eight time steps transient trajectory to ZLS (Fig. 3b). In the first part of the experiment, Carbachol (20 $\mu$M), a modulator of cholinergic receptors, was added to the network. As expected \cite{18}, Carbachol suppresses the synchronous mode of the network. The stability of ZLS with transmission delay of 0.5 s was estimated by observing the activity periods in 20 trials. Specifically, we estimated the "survival duration" of reverberating patterns by measuring the time to first occurrence where no evoked spikes were detected in all four electrodes. The average survival duration is 43.7 s in the presence of Carbachol (Fig. 3c). Figure 3d (left) shows that the above stability is maintained albeit significant intermittent responsiveness. The other extreme pharmacological manipulation used here is addition of Bicuculline (5 $\mu$M), leading to a suppression of total network activity, amplification of correlated bursting activity interleaved with long (around 20 s on average) periods of practically complete quiescence. The stability of ZLS with transmission delay of 0.5 s was estimated as explained above and shown in Fig. 3c. In this case, the bursts may block the synthetic reverberating pattern completely, leading to 6.95 s average survival duration. As a result, most trials terminate before reaching the fully blown reverberating pattern of activity. An atypical trial, where a reverberating pattern is arrived at, is shown in Fig. 3d (right). We believe that the median survival duration with Bicuculline (1.5 s; Fig. 3c) on the background of the long average periods of complete quiescence (20 s), reflects the emergence of burst activity simultaneously with the stimulation activity of the circuit every 0.5 s.

\begin{figure}[h]
\begin{center}
\includegraphics[width=0.8\textwidth]{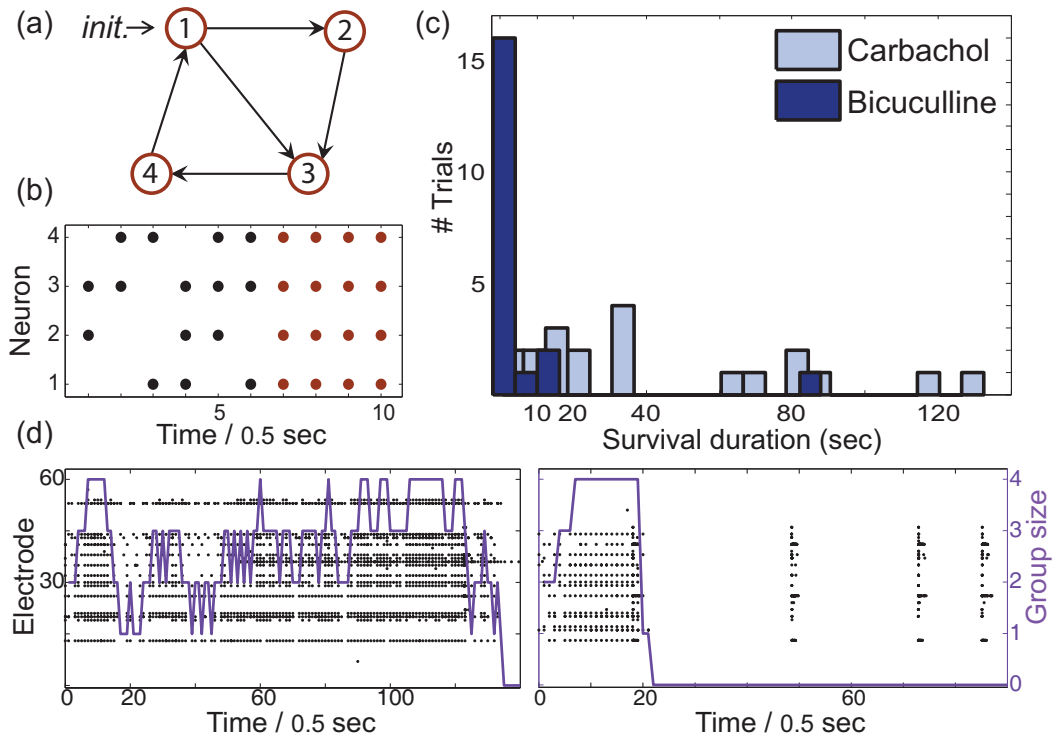}
\end{center}
\caption{ Robustness of synthetic reverberating pattern to network spontaneous and burst activities. (a) The smallest homogeneous directed circuit leading to ZLS, four electrodes with 3 and 4 directed loops. (b) The transient trajectory to ZLS where a stimulation is given to neuron labelled 1. (c) Histogram of sessions of 20 trials with transmission delay of 0.5 s. The average/median survival duration  with Carbachol (light blue) is 43.7/31 s and with Bicuculline (dark blue) 6.93/1.5 s. (d) Raster plot of the activity of the sixty electrodes (black) and the sum of the intermittent responsiveness of the four electrodes (violet). A typical trial with Carbachol leading to significant intermittent responsiveness (left), and an atypical trial with Bicuculline where a burst activity leads to a sudden death of the synthetic reverberating pattern (right).}
\end{figure}

\section{Complex external stimuli}
The interplay between the GCD circuit loops and the selected synthetic reverberating pattern initiated by stimulation to one electrode only, calls for generalization to stimulation given simultaneously to several electrodes. Fig. 4a demonstrates that a circuit consisting of 3 and 6 directed loops, in which the initial stimulation is given to electrodes labelled 1, 3 and 5 and the transmission delay is 2 s, results in ZLS, in spite of the fact that the GCD(3,6)=3. This demonstration indicates that the number of groups is equal to the GCD of the spatial circuit loops and the periodicity of the initial stimulation \cite{16}, i.e. GCD(2,3,6)=1 (Fig. 4a). One implication of the above spatiotemporally generalized GCD rule for the number of reverberating groups, is that spontaneous or network-induced fluctuations in the size or identity of zero-lag groups might lead to a phenomenon of switching between synthetic reverberating patterns: unexpected evoked (or omitted) spikes may be thought of as equivalent to initial stimulation through several electrodes simultaneously. Indeed, we observed such transitions between long-lasting patterns (Fig. 4b).

\section{Transient trajectory}
Since neural phenomena often occur on very short time scales, there seems to be an inherent temporal gap between the relatively long transient trajectory to steady state pattern, and the postulate that computing and functionality of the network are based on reverberating patterns. Moreover, it is clear that the igniting stimuli cannot be inferred from the reverberating pattern as exemplified by the six neuron circuit (Fig. 4c) where all 63 different stimulations lead to ZLS. An alternative computational view is akin to order-based representation \cite{15,20,21,22}, such that an inference takes place in the transient trajectory of activity to the reverberating pattern, where different igniting stimuli result in different or only partially overlapped transient trajectories (Fig. 4d-e). Under this computational framework, the GCD based steady state pattern stands mainly as a marker for the end of transient trajectory activity. The partial overlap between transient trajectories is otherwise unavoidable (each one of the stimulations points directly to the steady state pattern), however it enables a probabilistic inference which is accumulated dynamically along the trajectory, e.g. identifying a number or a subset of possible stimulated neurons (Fig. 4c).

\section{Concluding remarks}
There are several constraints, imposed by the experimental setup, that limit our capacity to generalize the above results to the brain. For instance, each node of a synthetic reverberating pattern in our experimental design is represented by a single neuron whose activation leads to a very strong excitation of the next neuron after an extremely long delay. This is in sharp contrast to the size of a functional cortical entity. Moreover, the present setup does not allow for mimicking a circuit with many members in each node
and fluctuating delays among pairs of connecting neurons,
and ignored spikes that happened during the delays.

In an extensive set of simulation studies at the population dynamics level we demonstrate that when the above constraints are relaxed, the general picture that emerges from our experimental analysis remains valid, e.g. Fig. 5. Every neural cell was simulated using the well known Hodgkin Huxley model \cite{23} with similar selected parameters as in \cite{16}. Synaptic background noise was simulated by a balance of input from 800 excitatory neurons firing at about 1-3Hz, and 200 inhibitory neurons firing at about 50-100 Hz. The free parameters were set in a way such that a single cell with no synaptic or external input fired randomly at about 5-7 Hz and the cell activity was noisy around the resting potential with a variance of about 5 mV. The connection between excitatory neurons belonging to different nodes was elected with probability $0.2$ and time delays among connecting neurons are taken from a uniform distribution between [29,31]ms.
Our numerical explorations indicate that when a single circuit node is a population of neurons, the signal to noise ratio of the  steady state is enhanced with the population sizes, and the circuit activity becomes less sensitive to background fluctuations. Nevertheless, the suitability of our neuronal circuits to describe irregular firing patterns of cerebral cortical area \cite{24,25} needs a further
investigation and might requires averaging out the regular firing by looking at long recording times.

\begin{figure}[h]
\begin{center}
\includegraphics[width=0.8\textwidth]{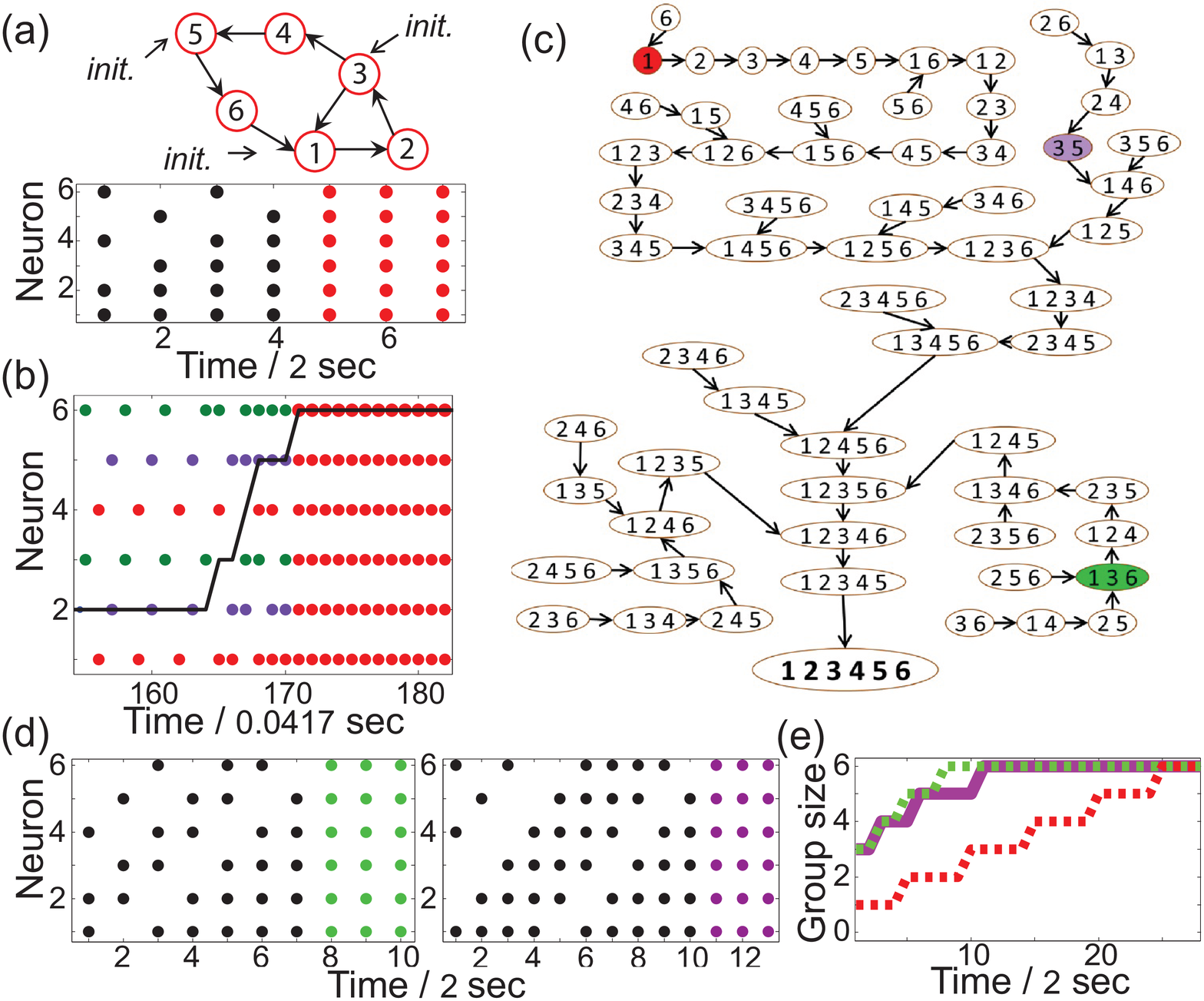}
\end{center}
\caption{ Steady state reverberating patterns reflect features of the circuit's igniting stimuli. (a) 3 and 6 directed loops where initial stimulation is given to electrodes labelled by 1, 3 and 5. The periodicity of the initial stimulation is 2 and the spatiotemporally generalized GCD(2,3,6)=1 leading to ZLS. (b) Spike trains of the circuit 1d with 3 periodical modes and with transmission delay of 1/24 s. The network noise leads to a transition between two long-lasting modes, from three modes to one mode during [165,170] time steps. At the steady state mode synchronized neurons are represented by the same colour and the black line stands for the sums up the number of simultaneously active electrodes. (c) An exhaustive enumeration of transients to ZLS of all the 63 ($2^6-1$) possible simultaneous stimulations of Fig. 1b. (d) Experimental result of the transient to ZLS with initial simultaneous stimulations to electrodes labelled (1,3,6) (green) and (3,5) (violet) with transmission delay of 2 s. (e) Sums up the number of electrodes with simultaneous activity for initial stimulation to electrodes labelled by (1,3,6) (dashed-green), (3,5) (violet) and (1) (dashed-red).}
\end{figure}

\begin{figure}[h]
\begin{center}
\includegraphics[width=0.6\textwidth]{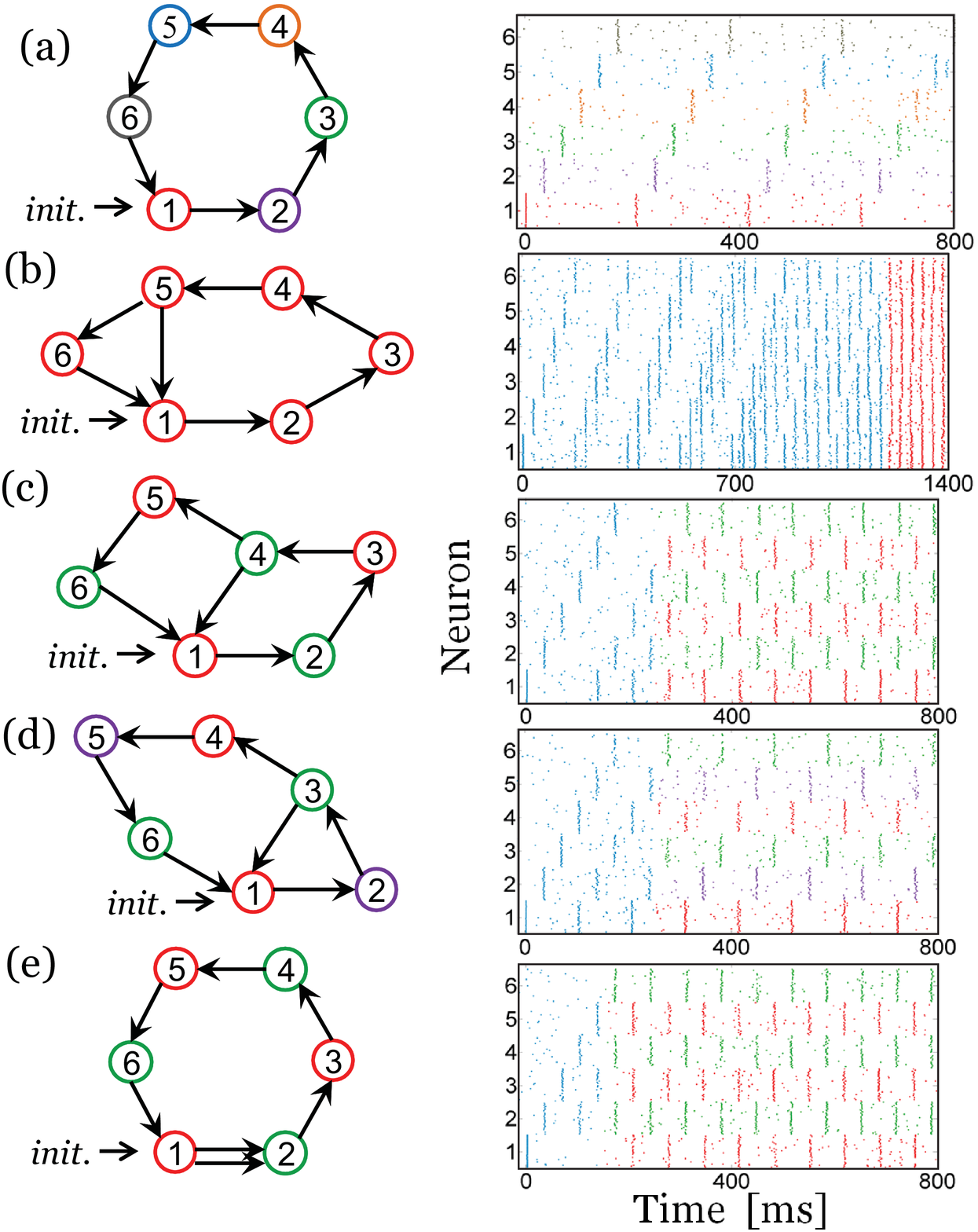}
\end{center}
\caption{ Raster diagram of the firing activity of the neurons in the population dynamics were the first row  is identical to  the first row of  Fig. 1. Each circuit node is represented by $30$ neurons and the entire six node circuit consists of $180$ neurons.  Simulations parameters are similar to the parameters used in \cite{16}.}
\end{figure}

\section{Materials and methods}
\subsection{Cell preparation}
Cortical neurons were obtained from newborn rats (Sprague- Dawley) within 24 h after birth using mechanical and enzymatic procedures described in earlier studies \cite{9,10,11,12,13,14,15}.  Rats were euthanized by $CO_2$  treatment according to protocols approved by the National Institutes of Health. The neurons were plated directly onto substrate-integrated multi-electrode arrays and allowed to develop functionally and structurally mature networks over a time period of 2-3 weeks. The number of plated neurons in a typical network is of the order of 1,300,000, covering an area of about 380 \begin{math}
mm^2 \end{math}. The preparations were bathed in MEM supplemented with heat-inactivated horse serum (5\%), glutamine (0.5 mM), glucose (20 mM), and gentamicin (10 g/ml), and maintained in an atmosphere of $37^0$C, 5\% $CO_2$, and 95\% air in an incubator as well as during the recording phases. The experiments described in Figures 1, 2 and 4 were conducted in the standard growth medium, supplemented with 50 \begin{math} \mu M \end{math} APV (amino-5-phosphonovaleric acid), 6.25 \begin{math}
\mu M \end{math} CNQX (6-cyano-7-nitroquinoxaline-2,3-dione), and 3.125  \begin{math} \mu M \end{math} Bicuculline; this cocktail of synaptic blockers made the spontaneous network activity sparse. Up-regulation and down-regulation of network spontaneous activity (Fig. 3) was achieved by addition of 20  \begin{math} \mu M \end{math} Carbachol and 5 \begin{math} \mu M \end{math} Bicuculline, respectively. In every event of pharmacological manipulation, at least half an hour was allowed for stabilization of the effect.
\subsection{Measurements and stimulation}
An array of 60 Ti/Au/TiN extracellular electrodes, 30 \begin{math} \mu m \end{math} in diameter, and spaced either 500 or 200 \begin{math} \mu m \end{math} from each other (MultiChannelSystems, Reutlingen, Germany) were used. The insulation layer (silicon nitride) was pre-treated with polyethyleneimine (Sigma, 0.01\% in 0.1 M Borate buffer solution). A commercial amplifier (MEA-1060-inv-BC, MCS, Reutlingen, Germany) with frequency limits of 150-3,000 Hz and a gain of $\times$1,024 was used. Mono-phasic square voltage pulses (200 \begin{math} \mu s \end{math}, 100-900 mV) were applied through extracellular electrodes. Data was digitized using data acquisition board (PD2-MF-64-3M/12H, UEI, Walpole, MA, USA). Each channel was sampled at a frequency of 16 K sample/s. Action potentials were detected on-line by threshold crossing. Data processing and conditioned stimulation were performed by using Simulink (The Mathworks, Natick, MA, USA) based xPC target application \cite{12}.

\subsection{Cell selection}
Each circuit node was represented by a stimulation source (source electrode) and a target for the stimulation- the recorded electrode (target electrode). The stimulation electrodes (source and target) were selected as the ones that evoked well isolated and well formed spikes and reliable response. This examination was done with stimulus intensity of 800 mV, after 30 repetitions at a frequency of 5 Hz. After the electrodes selection the experiment began.

\subsection{Stimulation Control}
 A circuit is formed by 6 or 4 nodes with programmable identical delays between nodes. A given node response was defined as a spike occurring within a time window of the observed neuronal response latency $\pm$2 ms following the electrical stimulus. The activity of all target electrodes for each circuit stimulation was collected and entailed stimuli were delivered in accordance to the adjacency matrix.

\acknowledgments
The authors thank Vladimir and Elleonora Lyakhov for invaluable technical assistance and Reut Timor for simulation assistant. The research leading to these results has received funding from the European Union Seventh Framework Program FP7 under grant agreement 269459 and grant of the Ministry of Science and Technology of the State of Israel and MATERA grant agreement 3-7878.


\end{document}